\documentclass[nolinenumbers]{aastex631}

\begin{document}

\title{Three types of solar coronal rain during magnetic reconnection between open and closed magnetic structures}

\author{Fangfang Qiao}
\affiliation{School of Earth and Space Sciences, Peking University, Beijing 100871, People's Republic of China}

\author[0000-0001-5776-056X]{Leping Li}
\affiliation{National Astronomical Observatories, Chinese Academy of Sciences, Beijing 100101, People's Republic of China; \textit{Email: \href{mailto:lepingli@nao.cas.cn}{lepingli@nao.cas.cn}}}
\affiliation{Key Laboratory of Solar Activity and Space Weather, National Space Science Center, Chinese Academy of Sciences, Beijing 100190, People's Republic of China}
\affiliation{University of Chinese Academy of Sciences, Beijing 100049, People's Republic of China}

\author[0000-0002-1369-1758]{Hui Tian}
\affiliation{School of Earth and Space Sciences, Peking University, Beijing 100871, People's Republic of China}
\affiliation{Key Laboratory of Solar Activity and Space Weather, National Space Science Center, Chinese Academy of Sciences, Beijing 100190, People's Republic of China}

\author[0000-0003-4804-5673]{Zhenyong Hou}
\affiliation{School of Earth and Space Sciences, Peking University, Beijing 100871, People's Republic of China}

\author[0000-0001-5705-661X]{Hongqiang Song}
\affiliation{Shandong Provincial Key Laboratory of Optical Astronomy and Solar-Terrestrial Environment, and Institute of Space Sciences, Shandong University, Weihai, Shandong 264209, People's Republic of China}

\author[0000-0001-8950-3875]{Kaifan Ji}
\affiliation{Yunnan Observatories, Chinese Academy of Sciences, Kunming 650216, People's Republic of China}

\author[0000-0001-5657-7587]{Zheng Sun}
\affiliation{School of Earth and Space Sciences, Peking University, Beijing 100871, People's Republic of China}


\begin{abstract}

Coronal rain (CR) is a crucial part of the mass cycle between the corona and chromosphere. 
It includes the flare-driven CR and two types of quiescent CR separately along the non-flaring active region closed loops and along the open structures, labeled as types I, II, and III CR, respectively.
Among them, types I and III CR are generally associated with magnetic reconnection. 
In this study, employing data taken by the Solar Dynamics Observatory (SDO) and the Solar Upper Transition Region Imager (SUTRI) on 2022 October 11, we report three types of CR during an interchange reconnection between open and closed magnetic filed structures above the southeastern solar limb. 
The open and closed structures converge, with the formation of current sheet at the interface, and reconnect. 
The newly-formed closed and open structures then recede from the reconnection region. 
During the reconnection, coronal condensation occurs along the reconnecting closed loops, and falls toward the solar surface along both loop legs as the type II CR.
Subsequently, condensation happens in the newly-formed closed loops, and moves down toward the solar surface along both loop legs as the type I CR.
Magnetic dips of the reconnecting open structures form during the reconnection.
In the dips, condensation occurs, and propagates along the open structures toward the solar surface as the type III CR. 
Our results suggest that the reconnection rate may be crucial for the formation of types I and III CR during the reconnection.

\end{abstract}

\keywords{Solar magnetic reconnection -- Plasma physics -- Solar ultraviolet radiation -- Solar corona -- Solar magnetic fields}

\section{Introduction} \label{sec:intro}

In two-dimensional models, magnetic reconnection occurs at an X-point, where oppositely directed magnetic field lines converge and undergo reconnection, rapidly releasing magnetic energy that is converted into kinetic and thermal energy through the acceleration or heating of charged particles \citep{RevModPhys.82.603, antolin2012disk, 2020RSPSA.47690867N}. 
It reconfigures the topology of magnetic field, and plays an elemental role in the magnetized plasma systems throughout the universe including the planetary magnetospheres, magnetars, and accretion disks around black holes \citep{MagneticReconnection}. 
Numerous theoretical concepts of reconnection have been undertaken to explain various features in solar physics, such as the flares, filament eruptions, jets, and coronal mass ejections \citep{shibata1998evidence, lin2000effects, 2021RSPSA.47700217S}.

In-situ measurements are not feasible in the extremely hot solar atmosphere. 
Fortunately, the magnetic field pressure dominates the plasma pressure (low plasma beta) in the solar corona and the magnetic flux is frozen into the highly conductive coronal plasma \citep{aschwanden2006physics, 2014masu.book.....P}. 
The emitting plasma within coronal loops not only outlines the magnetic field's geometry but also reflects changes in field connectivity. 
It enables observations through remote sensing of emission across the entire electromagnetic spectrum from radio to X-rays and gamma rays, and can serve as indirect indicators of magnetic field topology and evolution \citep{2013NatPh...9..489S, 2016NatPh..12..847L}.

So far, many signatures of reconnection have been reported using remote sensing observations. 
These include the reconnection inflows \citep{2001ApJ...546L..69Y, 2005ApJ...622.1251L, 2009ApJ...703..877L}, current sheets \citep{2008ApJ...686.1372C, 2010ApJ...723L..28L, 2012SoPh..276..261S, 2016NatPh..12..847L, 2016ApJ...829L..33L, 2018ApJ...866...64C, 2019ApJ...879...74C, 2019ApJ...874..146H, 2021ApJ...915...39H}, reconnection outflows \citep{takasao2011simultaneous, 2014ApJ...797L..14T, chen2016tether, li2016bi}, plasmoid ejections \citep{2012PhRvX...2b1015S, 2013A&A...557A.115K, 2019A&A...628A...8P, 2020A&A...633A.121X}, loop-top hard X-ray sources \citep{1994Natur.371..495M, 2013NatPh...9..489S}, supra-arcade downflows \citep{2000SoPh..195..381M, 2003SoPh..217..267I, 2016ApJ...829L..33L, 2019PhRvL.123c5102S, 2021Innov...200083S}, cusp-shaped post-flare loops \citep{1992PASJ...44L..63T, 2018ApJ...853L..18Y}, and coronal structural reconfigurations \citep{2013ApJ...776...57Z, 2014A&A...570A..93L, 2018ApJ...864L...4L, 2018ApJ...868L..33L, 2019ApJ...884...34L, 2018ApJ...863L..22K}. 
However, reconnection events with clear structural evolution and evident complete process are still relatively rare in observations.

Over the past fifty years, dozens of extreme ultraviolet (EUV) imagers and spectrographs have been deployed into space for solar observation purposes. 
Significant advancements have been made in terms of spatial resolution and effective area, enhancing our ability to study the Sun. 
The Atmospheric Imaging Assembly \citep[AIA;][]{lemen2012atmospheric} onboard the Solar Dynamics Observatory \citep[SDO;][]{pesnell2012solar} has been taking full-disk solar images in ten passbands including seven EUV passbands with a similar resolution. The Solar Upper Transition Region Imager (SUTRI), a narrow-band Ne\,{\sc vii} 46.5 nm imager on board the Space Advanced Technology demonstration satellite (SATech-01), takes images of the plasma with a characteristic temperature of $\sim$0.5 MK \citep{tian2017probing, bai2023solar}. 

In the solar atmosphere, there are numerous fascinating phenomena, e.g., flares and filament eruptions. 
Among them, coronal rain (CR) is a particularly well-known and frequently observed phenomenon.
First discovered in the 1970s \citep{1970PASJ...22..405K,leroy1972emissions}, CR plays an essential role in the mass cycle between the hot, tenuous corona and the cool, dense chromosphere \citep{2018ApJ...864L...4L, 2019ApJ...884...34L,2019A&A...630A.123K, li2020relation, li2021disk}. It is observationally characterized by the formation of cool, dense, and blob-like material within minutes in the hot corona, which subsequently descends along curved, loop-like trajectories to the solar surface \citep{2012ApJ...745..152A}. 
CR has been extensively observed both off-limb and on-disk, with its best observations typically achieved off-limb. 
Here, it appears in emission in the chromospheric and transition region lines, exhibiting high contrast against the dark background \citep{2020PPCF...62a4016A}. 
CR is a multithermal phenomenon, emitting in the chromospheric lines such as H$ \alpha $, Ca\,{\sc ii h}, and Mg\,{\sc ii k}, as well as in the transition region lines, e.g., Si {\sc iv}, with temperatures ranging from a few thousand to several hundred thousand Kelvin. 
Observations provide a comprehensive view of the multithermal coronal plasma, where emission of different wavelengths is cospatial \citep{2015ApJ...806...81A}. 
As for the observed properties, the lengths of CR blobs range from a few hundred km to 1-20 Mm, with peak values of 0.7–1 Mm \citep{2012ApJ...745..152A}, and widths range from 150 to 300 km \citep{2015ApJ...806...81A, 2023A&A...676A.112A, 2022ApJ...926..216L, 2023ApJ...950..171S}.
The material with densities of 10$^{10}$–10$^{11}$ cm$^{-3}$ falls to the solar surface with speeds in the range from a few 10 to a few 100 km s$^{-1}$, with the peak at 80–100 km s$^{-1}$ \citep{2001SoPh..198..325S, 2009RAA.....9.1368Z, 2012ApJ...745..152A, 2015ApJ...806...81A}.

While CR is usually seen falling towards the solar surface along the legs of a loop, there are instances of upward flow on one leg followed by downward flow on the other (partial siphon flow), as well as changes of direction (upward/downward) along the same leg \citep{2010ApJ...716..154A,2022ApJ...926..216L,2023ApJ...950..171S}. 
A particular aspect of CR dynamics is its downward acceleration being less than the free-fall acceleration (flow along magnetic field lines). 
Several studies, using various imaging and spectroscopic instruments capable of capturing the complete velocity vector, not just the apparent plane-of-sky component, have reported downward acceleration of about or less than 0.1 km\,s$^{-2}$ \citep{2014SoPh..289.4117A, 2012ApJ...745..152A, 2001SoPh..198..325S, 2004A&A...415.1141D, 2016ApJ...827...39K, 2017A&A...598A..57V}. 
Comparing with the effective solar gravitational acceleration, this acceleration is notably smaller. 
By restructuring the downstream of the CR, the gas pressure may result in the observed smaller acceleration \citep{2014ApJ...784...21O}.
Additionally, when transverse magnetohydrodynamic (MHD) waves are present, the transverse MHD waves' suspension force does not play a major role \citep{2016ApJ...827...39K,2011ApJ...736..121A}. 
While the heating mechanism of loops is expected to produce sustained gas pressure changes within the loops, the formation of a dense clump in a rarefied medium like the corona generates a leading gas pressure front that propagates along the magnetic field lines, enhancing the gas pressure at its rear and rapidly counteracting the downward acceleration \citep{2010ApJ...716..154A,2017A&A...602A..23K,2014ApJ...784...21O,2016ApJ...818..128O}. 
Recent studies have also found that the strong gas pressure gradient above and below the dense group acts as an antigravity levitation force that influences the dynamics of CR \citep{2022ApJ...926..216L}.

Thermal instability is widely recognized as a significant mechanism for the formation of condensed material in CR and prominences/filaments \citep{1953ApJ...117..431P,1965ApJ...142..531F,2019A&A...624A..96C,2020A&A...636A.112C}.
The distinction between the two activities is largely influenced by their local magnetic field configurations. 
Condensation formed in an arched loop may descend along the loop legs and manifest as CR within a short period, while condensation in a prominence  may linger for hours to days, supported by magnetic field depressions \citep{1994ApJ...420L..41A,2021A&A...649A.142A,2021A&A...646A.134J}. 
Thermal instabilities occur when small primary perturbations disrupt the density and temperature equilibrium of the homogeneous coronal plasma, which is balanced between heating and cooling \citep{2019SoPh..294..173K}. 
It is typically assumed that chromospheric plasma evaporates due to heating at the loop footpoints, filling the loop and making it denser and hotter \citep{1999ApJ...512..985A,2001ApJ...560.1035A,2001ApJ...553L..85K,2003A&A...411..605M,2004A&A...424..289M,2005A&A...436.1067M,2011ApJ...737...27X}. 
As a result, radiative losses become more efficient, and loops begin to cool when the heat transfer from heating is insufficient to balance the radiative losses. 
Since the radiative loss function is negatively correlated with temperature in the coronal temperature range, a decrease in temperature further increases radiation, leading to catastrophic cooling. 
The decrease in local temperature in the corona also reduces pressure, causing surrounding material to converge, forming localized material concentration regions. 
This dense matter is effectively cooled by radiation until the plasma becomes more optically thick and the cooling rate decreases. 
Modern views also consider the possibility that this thermal equilibrium may never be achieved, leading to a thermal non-equilibrium cycle \citep{2020PPCF...62a4016A}.

CR is basically classified into two flavours based on its relation with the flare: the flare-driven CR, labeled as the type I CR, and the quiescent CR \citep{2020PPCF...62a4016A,2021RAA....21..255L}. 
Additionally, the quiescent CR is sorted into two kinds  according to the magnetic field structures where it forms \citep{li2020relation,li2021disk}. 
One takes place along the non-flaring active region (AR) closed loops \citep{2001SoPh..198..325S, 2010ApJ...716..154A}, labeled as the type II CR, and the other occurs along the open magnetic structures in both quiet-Sun regions \citep{2018ApJ...864L...4L} and ARs \citep{2019ApJ...874L..33M, 2021RAA....21..255L}, labeled as the type III CR. 
For these three kinds of CR, different formation mechanisms have been widely investigated \citep{2003A&A...411..605M, 2004A&A...424..289M, 2014ApJ...797...36S, 2016ApJ...833..184S, 2018ApJ...864L...4L, 2019ApJ...884...34L, li2020relation, li2021disk, 2019ApJ...874L..33M, 2020PPCF...62a4016A}.

The type I CR often occurs in post-flare loops \citep{2003ApJ...586.1417B,2014ApJ...797...36S,2016ApJ...833..184S,2017ApJ...842...15L}. 
It has been suggested that the condensed material of this type of CR forms because the directed non-thermal particles and/or thermal conduction fronts produced during the flare rapidly heat the chromospheric material, causing the heated chromospheric material to evaporate to fill the post-flare loops, increasing their mass \citep{2009ApJ...690..347L, 2015ApJ...811..139T, 2016ApJ...827...27Z}. 
At the top of the post-flare loops, thermal instability occurs when the radiative loss exceeds the heating input \citep{1953ApJ...117..431P,1965ApJ...142..531F}. 
As a result, the thermally evaporating plasma rapidly cools and condenses. 
The condensation then falls toward the solar surface, under the effect of gravity, along both legs of post-flare loops as the type I CR.
Here, the electron beam heating alone cannot directly produce the CR in post-flare loops \citep{2020ApJ...890..100R}.

For the  formation of condensation in the type II CR \citep{2003A&A...411..605M, 2004A&A...424..289M}, heating events believed to be concentrated at/near the loop footpoints lead to chromospheric material evaporation and direct mass ejections into the AR closed loops \citep{patsourakos2006nonthermal, 2015A&A...583A.109L}. 
Consequently, the loops quickly become hotter and denser, and then rapidly cool and condense due to thermal instability within a loop in a state of thermal non-equilibrium \citep{2010ApJ...716..154A, 2020PPCF...62a4016A}.
The condensation falls toward the solar surface along both legs of the loops as the type II CR under the effect of gravity.
Moreover, along loops, the formation of the type II CR, possibly triggered by impulsive heating associated with reconnection, has also been observed \citep{2019A&A...630A.123K}. 
However, similar event showing the type II CR associated with reconnection is rarely reported.
Unlike the type I CR, the type II CR is a (quasi-) periodic phenomenon with intervals of a few hours \citep{antolin2012disk,2018ApJ...853..176A,2016ApJ...827...39K,2020A&A...633A..11F}. 
This periodicity of stationary CR may be attributed to thermal non-equilibrium \citep{2018ApJ...853..176A,2020A&A...633A..11F}.

A recent model was proposed to elucidate the generation of condensation in the quiescent CR along open field lines, i.e., the type III CR, facilitated by interchange reconnection \citep{2018ApJ...864L...4L}. 
In this model, the curved higher-lying open structures descend toward the solar surface and undergo interchange reconnection with the lower-lying closed loops, leading to the creation of a magnetic dip in the former. 
Subsequently, the newly-formed open structures and closed loops appear and recede from the reconnection region. In the higher-lying open structures, the coronal plasma around the dip converges into the dip, leading to an increase in plasma density in the dip.
This density enhancement triggers thermal instability, leading to the cooling and condensation of hot coronal plasma within the dip. 
A prominence is thus formed \citep{1991ApJ...378..372A,yan2015formation,2020RAA....20..166C}, facilitating a speedup of the reconnection.
Due to the successive reconnection, under the effect of gravity, the condensed material falls to the solar surface along the legs of the newly-reconnected closed loops and the higher-lying open structures as the type III CR \citep{2018ApJ...868L..33L}.

The types I and III CR are generally associated with reconnection. 
Both of them are separately observed during the interchange reconnection between open and closed structures \citep[e.g.,][]{2016ApJ...830L...4S, 2018ApJ...864L...4L}.
However, it is still not clear under what conditions the types I and III CR form during the interchange reconnection.
In this study, we utilize the SDO and SUTRI data to analyze an interchange reconnection event with a distinctive structure and clear evolution. 
Notably, for the first time, three types of CR are observed in the reconnection event. 
The conditions under which the types I and III CR form are then investigated.
The observations, results, and summary and discussions are described in Sections\,\ref{sec:Observations}, \ref{sec:results}, and \ref{sec:summary}, respectively.

\section{Observations} \label{sec:Observations}
The SDO/AIA comprises a set of normal-incidence imaging telescopes that capture solar atmospheric images across ten wavelength bands, including seven EUV channels and three ultraviolet (UV) channels \citep{lemen2012atmospheric}.
The AIA EUV and UV images are taken with a time cadence of 12 and 24\,s, respectively, and have a spatial sampling of 0.6\arcsec\,pixel$^{-1}$.
Each AIA channel corresponds to plasma at different temperatures, e.g., the 94\,\AA~channel captures plasma at $\sim$7.2 MK (Fe\,{\sc xviii}), the 335\,\AA~channel captures plasma at $\sim$2.5 MK (Fe\,{\sc xvi}), the 211\,\AA~channel captures plasma at $\sim$1.9 MK (Fe\,{\sc xiv}), the 193\,\AA~channel captures plasma at $\sim$1.5 MK (Fe\,{\sc xii}), the 171\,\AA~channel captures plasma at $\sim$0.9 MK (Fe\,{\sc ix}), the 131\,\AA~channel captures plasma at $\sim$0.6 MK (Fe\,{\sc viii}) and $\sim$10 MK (Fe\,{\sc xxi}), and the 304\,\AA~channel captures plasma at $\sim$0.05 MK (He\,{\sc ii}).

The scientific objective of the science and technological demonstration payload SUTRI is to establish a connection between the lower solar atmosphere and the structure of the corona \citep{bai2023solar} . 
It operates in a Sun-synchronous orbit at an altitude of approximately 500 km and aims to enhance our understanding of various solar activities such as flares, jets, and coronal mass ejections. 
SUTRI primarily focuses on the Ne\,{\sc vii} 465\,\AA~spectral line, which corresponds to a characteristic temperature of $\sim$0.5 MK in the solar atmosphere \citep{tian2017probing}. 
With a filter width of about 30\,\AA, it achieves a moderate spatial resolution of $\sim$8\arcsec, allowing for full-surface solar atmospheric images. 

In this study, we employ all available SDO/AIA EUV, i.e., 94, 335, 211, 193, 171, 131, and 304\,\AA, images, and SUTRI 465\,\AA~images to conduct the comprehensive investigation of the interchange reconnection event on 2022 October 11. 
The alignment of AIA EUV and SUTRI 465\,\AA~images is achieved using a rapid subpixel image registration method \citep{2012IEEE, 2015RAA....15..569Y}. Data from SUTRI 465\,\AA~has been aligned by optimizing cross-correlation with AIA 304\,\AA~images, as they show the most similar characteristic features.

\section{Results} \label{sec:results}

Between 05:00 and 16:00 UT on October 11, 2022, a sustained interchange reconnection event occurs above the southeastern solar limb. 
By investigating the evolution of the region, where the event happens, in the following 10 days using AIA EUV images and line-of-sight (LOS) magnetograms of the Helioseismic and Magnetic Imager \citep[HMI;][]{2012SoPh..275..229S} onboard the SDO, we notice that the reconnection takes place in a large bipolar magnetic region with positive field, PF1, in the north and negative field, NF1, in the south. Moreover, there is another positive field, PF2, located to the south of the bipolar region.
At 00:00\,UT on 2022 October 13, about two days after the event, the magnetic fields PF1, NF1, and PF2 are located at the heliographic positions S15\,W60, S18\,W57, and S27\,W60, respectively.
We use SDO/AIA EUV and SUTRI 465\,\AA~images to study the magnetic structures and their evolution in the solar corona. 
We find that a set of open structures, denoted as L2, move downward, while a set of closed loops, labeled as L1, move upward (Figure\,\ref{fig1}). 
Here, the closed loops L1 connect the negative and positive fields NF1 and PF2, while the open structures L2 root in the positive field PF1.

These structures eventually converge and transform into new sets of closed loops, denoted as L3, and open structures, labeled as L4 (Figure\,\ref{fig1}). 
These newly-formed structures L3 and L4 then separate from each other. 
Details are presented in Section \ref{subsec:MR}. 
Moreover, abundant CR activities are observed in AIA 304\,\AA~and SUTRI 465\,\AA~images. 
We analyze the morphology and evolution of the CR activities and identify a total of three types of CR. 
This is the first time where all three types of CR have been observed in a single reconnection event; see details in Section\,\ref{subsec:coronal rain}.

\subsection{Interchange reconnection} \label{subsec:MR}

\begin{figure}[htbp]
    \centering
    \includegraphics[width=1.0\textwidth]{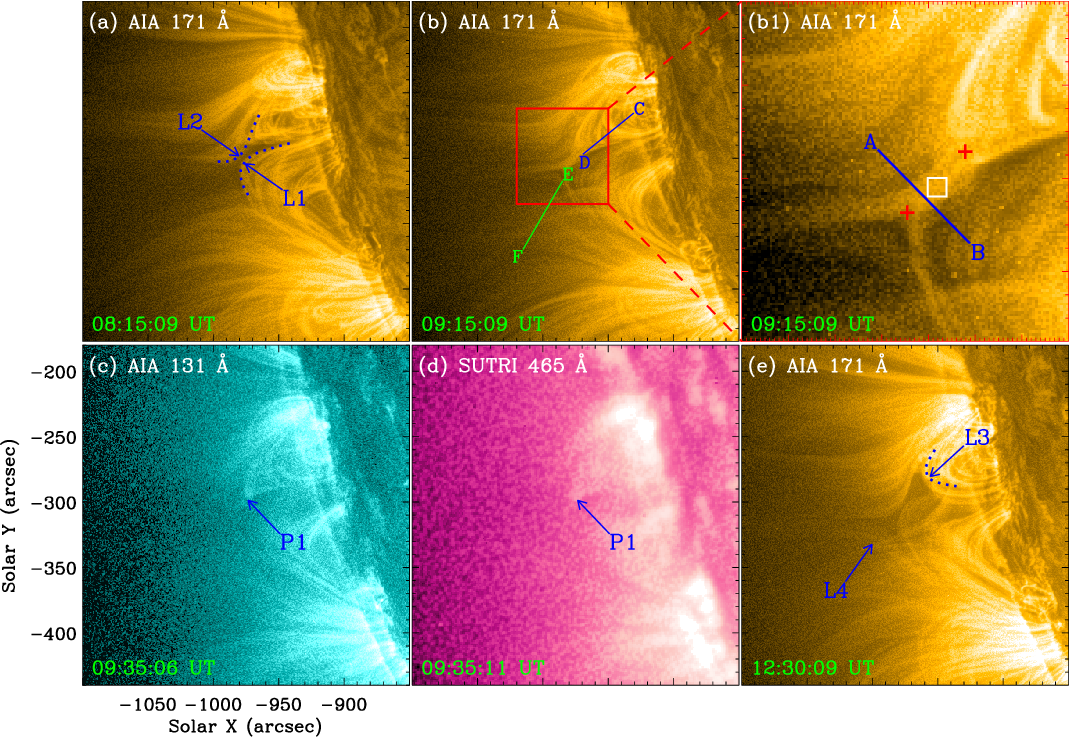}
    \caption{Magnetic reconnection between open and closed structures. 
    (a), (b), (b1) and (e) SDO/AIA 171 and (c) 131\,\AA, and (d) SUTRI 465\,\AA~images. 
    The blue dotted lines in (a) outline the closed and open structures L1 and L2. 
    The red rectangle in (b) indicates the field of view of (b1). 
    The blue line AB in (b1), and the blue and green lines CD and EF in (b) mark the positions for the time slices in Figures\,\ref{fig2}(a)-(c), respectively. 
    The white rectangle in (b1) encloses the region for the DEM curve in Figure\,\ref{fig2}(d). 
    Two red pluses in (b1) mark the positions between which the length of the current sheet is measured.
    The cusp-shaped structure, P1, is marked by the blue arrows in (c) and (d). 
    The blue dotted line in (e) outlines the newly-reconnected closed loops, L3. L4 in (e) represents the newly-reconnected open structures.
    An animation of the unannotated AIA images (panels (a) and (c)) is available.
    It covers $\sim$10\,hr, starting at 06:00 UT, with a time cadence of 1\,minute.
    See Section\,\ref{subsec:MR} for details.
    (An animation of this figure is available.)}
    \label{fig1}
\end{figure}

As depicted in Figure\,\ref{fig1}(a), two sets of coronal structures exhibit distinct saddle-shaped features in an AIA 171\,\AA~image at 08:15 UT.
The northern set corresponds to open structures L2, while the southern set represents closed loops L1. 
The two sets of structures are close to each other at the top, forming an X-shaped configuration; see Figure\,\ref{fig1}(a). 
Subsequently, at 09:15 UT, as illustrated in Figure\,\ref{fig1}(b), the two sets of structures converge and interact. 
At the interface, a distinct current sheet structure is observed.
In Figure\,\ref{fig1}(b), we enclose the region containing the current sheet structure with a red rectangle and provide a zoomed-in view in Figure\,\ref{fig1}(b1).
The current sheet has a length of $\sim$13\,Mm and a width of $\sim$5.3\,Mm. 
Here, we determine the length of the current sheet by measuring the distance between the two cusp-shaped structures located at its ends, as indicated by two red pluses in Figure\,\ref{fig1}(b1). To determine the current sheet width, we initially extract the intensity profile in the AIA 171\,\AA~passband perpendicular to the current sheet, e.g., along the direction AB in Figure\,\ref{fig1}(b1). The background emission is obtained by averaging the intensity around the current sheet, which is then subtracted from the intensity profile. Subsequently, we apply a single Gaussian fit to the residual intensity profile, and use the full width at half maximum (FWHM) of the Gaussian fit as the width.
The reconnection rate, the ratio of the width and the length of the current sheet, is thus $\sim$0.4.
As shown in Figure\,\ref{fig1}(e), new closed loops L3 and open structures L4 form and retract away from the reconnection region. 
Moreover, we observe a distinct cusp-shaped structure in AIA 211, 193, 171, and 131\,\AA, and SUTRI 465\,\AA~images, which is shown in Figures\,\ref{fig1}(c) and (d). This structure is, however, not evidently identified in the AIA high-temperature passbands, e.g., 94 and 335\,\AA.

\begin{figure}[htbp]
    \centering
    \includegraphics[width=1.0\textwidth]{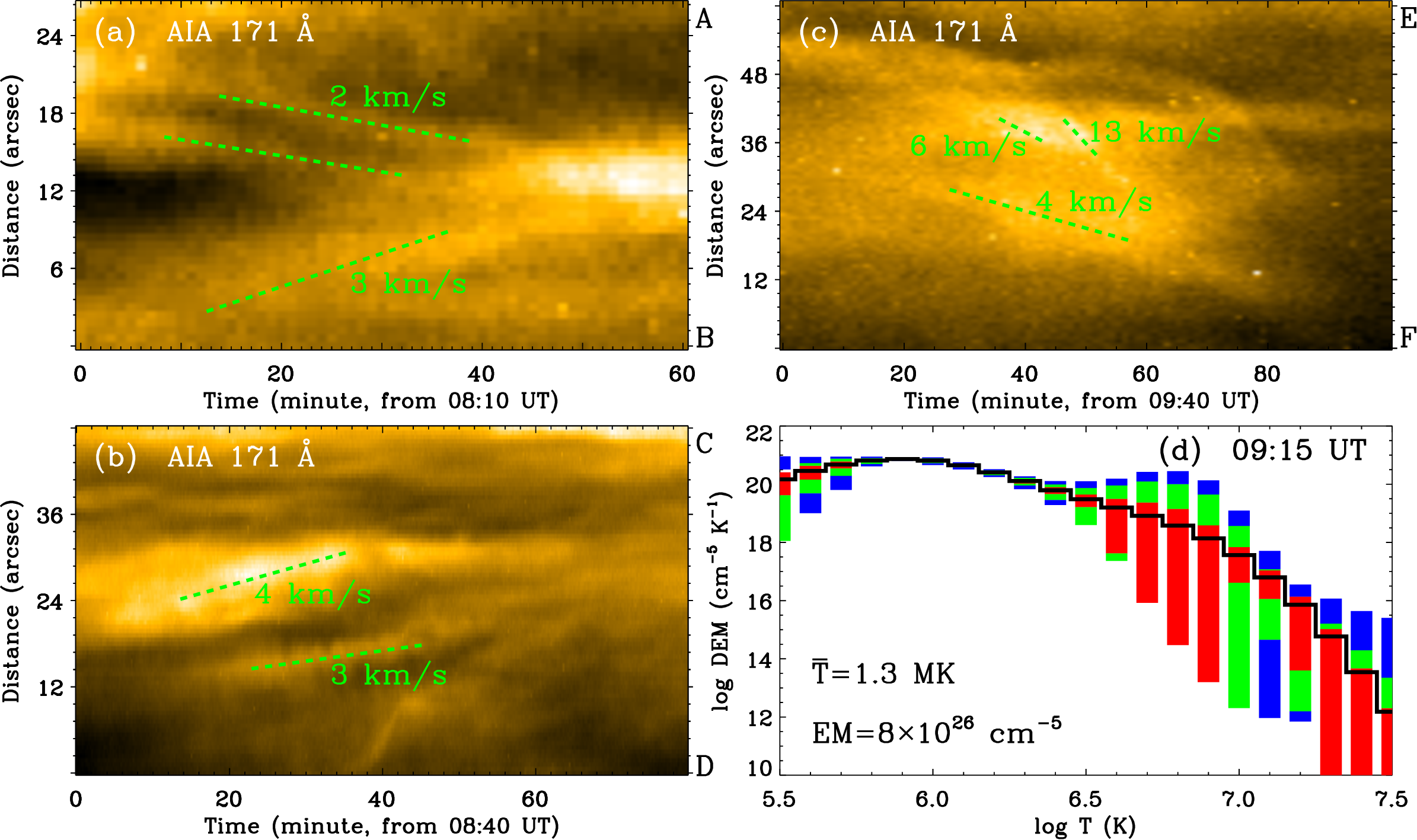}
    \caption{Temporal evolution of the reconnection between open and closed structures. 
    (a)–(c) Time slices of AIA 171\,\AA~images along the blue and green lines AB, CD, and EF in Figures\,\ref{fig1}(b1) and (b), respectively. 
    (d) DEM curve for the current sheet region enclosed by the white rectangle in Figure\,\ref{fig1}(b1). 
    The motions of structures L1-L4 are outlined by the green dotted lines in (a)–(c), with their respective speeds indicated by the numerical values. 
    The black line in (d) is the best-fit DEM distribution, while the red, green, and blue rectangles correspond to the regions containing 50\%, 51\%–80\%, and 81\%–95\% of the Monte Carlo solutions. 
    See Section\,\ref{subsec:MR} for details.}
    \label{fig2}
\end{figure}

To conduct a quantitative analysis of the reconnection inflows, we make a time slice of AIA 171 \AA~images along the line AB in Figure\,\ref{fig1}(b1), and display it in Figure\,\ref{fig2}(a). 
In Figure\,\ref{fig2}(a), bilateral inflows toward the reconnection region are clearly identified. 
The structures on both sides gradually move closer to the center, and the reconnection region becomes clearly visible with the evolution of time. 
As shown in Figure\,\ref{fig2}(a), the velocities of the inflows on the two sides are calculated to be $\sim$2 and $\sim$3 km s$^{-1}$, respectively. 
In the current event, the newly-formed open structures L4 and closed loops L3 move away from the reconnection region. 
The motion of the closed loops L3 is much more apparent when observing them in both the AIA 171 and 131\,\AA~passbands; see the online animated version of Figure\,\ref{fig1}. 
Time slices of AIA 171\,\AA~images along the blue and green lines CD and EF in Figure\,\ref{fig1}(b) are made, and shown in Figures\,\ref{fig2}(b) and (c), respectively.
The loops L3 move toward the solar surface with speeds of $\sim$3-4 km\,s$^{-1}$, and the open structures L4 move away from the solar surface with speeds of $\sim$4-13 km\,s$^{-1}$; see the green dotted lines in Figures\,\ref{fig2}(b) and (c).

For further elaboration of the thermodynamic characterization of the current sheet, we employ the differential emission measure (DEM) analysis using "xrt\_dem\_iterative2.pro" \citep{2004IAUS..223..321W, 2012ApJ...761...62C}.
It utilizes six AIA EUV channels, including 94, 335, 211, 193, 171, and 131\,\AA, to analyze the temperature and emission measure (EM) of the current sheet. 
We select the current sheet region enclosed by the white rectangle in Figure\,\ref{fig1}(b1) for calculating the DEM at 09:15\,UT. 
To eliminate the influence of the background, the same region at 05:50 UT, when no current sheet forms, can be regarded as the background emission area which needs to be removed. 
In each region, the digital number counts in each of the six AIA channels are adjusted for the exposure time and then averaged across all pixels to obtain a spatially averaged value.
The DEM curve obtained from our calculations is displayed in Figure\,\ref{fig2}(d). 
To estimate the uncertainties of the DEM, we compute 100 Monte Carlo realizations of the data, as illustrated in Figure\,\ref{fig2}(d).
The current sheet emission is well constrained in the temperature range of 5.6$\le\log T (K)\le$6.5, rather than the high temperatures.
Over this temperature range, we integrate the DEM, and obtain a DEM-weighted mean temperature of 1.3 MK and a total EM of 8$\times$10$^{26}$\,cm$^{-5}$.

Given that the corona is primarily composed of fully ionized particles due to its high temperatures. Using the EM, the electron number density (n$_{e}$) of the current sheet is estimated employing n$_{e}$=$\sqrt{\frac{EM}{D}}$, where D is the LOS depth of the current sheet. 
Assuming that the depth (D) equals the width (W) of the current sheet, then the density is n$_{e}$=$\sqrt{\frac{EM}{W}}$.
Employing EM=8$\times$10$^{26}$\,cm$^{-5}$ and W=5.3\,Mm, we estimate the density of current sheet, and obtain a value of 1.2$\times$10$^{9}$\,cm$^{-3}$.

\subsection{Three types of CR} \label{subsec:coronal rain}

During the interchange reconnection between the closed and open structures L1 and L2, numerous instances of CR phenomena have been observed.
After a thorough investigation, three known types of CR, i.e., the types I, II, and III CR, are identified \citep{2021RAA....21..255L}.

\subsubsection{The type I CR} \label{subsubsec:3.1}

\begin{figure}[htbp]
    \centering
    \includegraphics[width=1.0\textwidth]{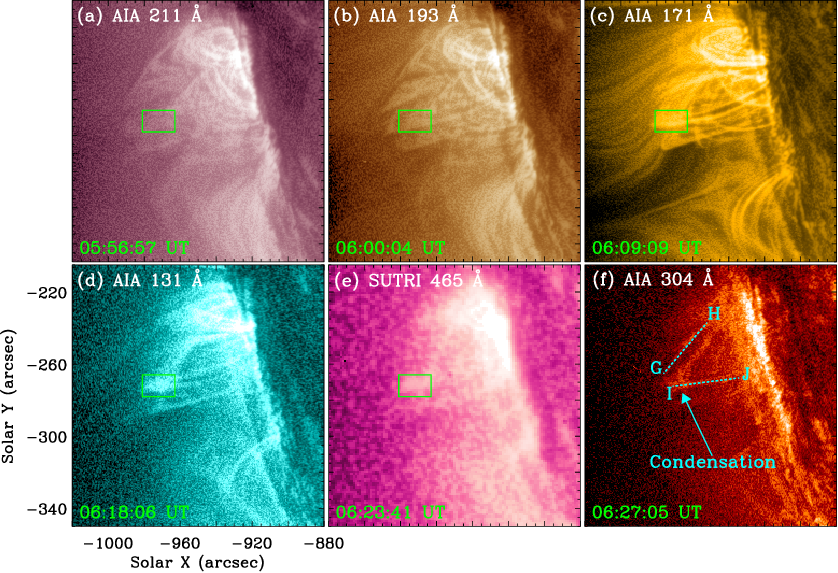}
    \caption{The type I CR. 
    (a) AIA 211, (b) 193, (c) 171, (d) 131 and (f) 304\,\AA, and (e) SUTRI 465\,\AA~images. 
    The green rectangles in (a)-(e) denote the region for the light curves in Figures\,\ref{fig7}(a1)-(a5). 
    In (f), the cyan dotted lines GH and IJ show the positions for the time slices in Figures\,\ref{fig6}(a1)-(a2). 
    An animation of the unannotated AIA and SUTRI images is available.
    It covers $\sim$6.2\,hr starting at 05:48\,UT, with a time cadence of 1\,minute.
    See Section\,\ref{subsubsec:3.1} for details.
    (An animation of this figure is available.)}
    \label{fig3}
\end{figure}

At 06:27 UT, the condensation is observed in AIA 304\,\AA~images, which show plasma at the characteristic temperature of $\sim$0.05 MK, accompanied by localized bright emission at the same positions in the AIA 211, 193, 171, 131\,\AA~and SUTRI 465\,\AA~images at different times; see Figures\,\ref{fig3}(a)–(f). 
Located at the top of the newly-reconnected loops L3, it is indicated by a cyan solid arrow in Figure\,\ref{fig3}(f). 
After a developmental period, this condensation begins to fall toward the solar surface along both legs of the loops L3; see the online animated version of Figure\,\ref{fig3}.
We make time slices of AIA 304\,\AA~images along the cyan dotted lines GH and IJ in Figure\,\ref{fig3}(f), and show them in Figures\,\ref{fig6}(a1) and (a2), respectively. 
Starting at $\sim$06:20 UT, a descending speed of $\sim$47 km\,s$^{-1}$, with an acceleration of $\sim$160 m\,s$^{-2}$, is observed along the direction GH of the condensation, as depicted by the green dotted lines in Figure\,\ref{fig6}(a1). 
The velocity of the condensation along another direction IJ is $\sim$24 km s$^{-1}$, with an acceleration of $\sim$80\,m\,s$^{-2}$. 
From Figures\,\ref{fig6}(a1) and (a2), we observe that the condensation initially accelerates during its descent but then maintains a roughly constant velocity. 
This behavior is likely due to the dominance of gravity downward along the magnetic field structure at first. 
Subsequently, this gravitational force may be balanced by the gas pressure gradient force, which acts opposite to gravity and is caused by downward compression.

We suggest that the interchange reconnection between the closed and open structures L1 and L2 leads to the evaporation of the chromospheric material, resulting in an increase in the mass load of the newly-formed loops L3.
This increase in mass causes thermal instability to occur at the top of the loops, allowing the hot evaporating plasma to cool down and condense rapidly. 
Due to the influence of gravity and gas pressure gradient, the condensation falls to the solar surface along the loops to form the CR. 
The formation mechanism of this type of CR is consistent with that of the type I CR, i.e., the flare-driven CR along post-flare loops.

The AIA EUV channels are generated by plasma radiation at various temperatures. 
Therefore, in order to better describe the process of thermal plasma cooling from higher temperatures, the light curves of different AIA  channels in the condensation region are calculated. 
The region of the type I CR where the light curves are computed is enclosed by the green rectangles in Figures\,\ref{fig3}(a)-(e). 
The normalized light curves of the AIA 211, 193, 171, 131 and 304\,\AA~and SUTRI 465\,\AA~channels are measured separately for two hours starting at 05:20 UT to obtain Figures\,\ref{fig7}(a1)-(a5). 
Here, the 131\,\AA~light curve shows plasma with the lower characteristic temperature of $\sim$0.6\,MK, rather than the higher characteristic temperature of $\sim$10 MK, as the newly-formed loops L3 are not detected in AIA 94 and 335\,\AA~images.
The plasma in the loops is hence heated up to $\sim$1.9\,MK, the characteristic temperature of AIA 211\,\AA~channel.

The moments when the peaks of the light curves appear are marked by the red vertical dotted lines in Figures\,\ref{fig7}(a1)-(a5). 
The 211, 193, 171, 131, and 304\,\AA~light curves peak at 05:57, 05:58, 06:09, 06:19, and 06:28 UT, respectively. 
This supports the cooling process of the plasma in the loops L3. 
The coronal plasma hence cools down from $\sim$1.9 MK (AIA 211\,\AA) to $\sim$1.5\,MK (AIA 193\,\AA) in 1 minute, to $\sim$0.9\,MK (AIA 171\,\AA) in 12 minutes, to $\sim$0.6\,MK (AIA 131\,\AA) in 22 minutes, and to $\sim$0.05\,MK (AIA 304\,\AA) in 31 minutes.
We can see that the SUTRI 465\,\AA~light curve; see the blue pluses in Figure\,\ref{fig7}(a4), exhibits a similar trend to the AIA 131\,\AA~light curve. 
This further supports that the loops L3 in AIA 131\,\AA~images show plasma at the lower characteristic temperature ($\sim$0.6\,MK) of the AIA 131\,\AA~channel.

\subsubsection{The type II CR} \label{subsubsec:3.2}

\begin{figure}[htbp]
    \centering
    \includegraphics[width=1.0\textwidth]{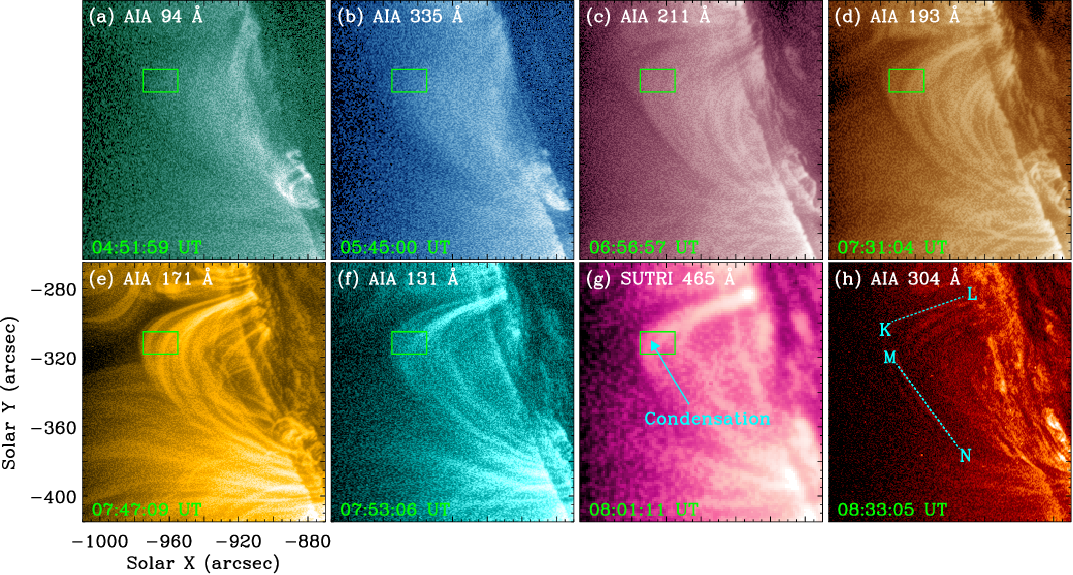}
    \caption{The type II CR. 
    (a) AIA 94, (b) 335, (c) 211, (d) 193, (e) 171, (f) 131, and (h) 304\,\AA, and (g) SUTRI 465\,\AA~images. 
    The green rectangles in (a)-(g) indicate the region for the light curves in Figures\,\ref{fig7}(b1)-(b7). 
    The cyan dotted lines KL and MN in (h) indicate the positions for the time slices in Figures\,\ref{fig6}(b1)-(b2). 
    An animation of the unannotated AIA and SUTRI images is available.
    It covers $\sim$3\,hr starting at 07:00\,UT, with a time cadence of 1\,minute.
    See Section\,\ref{subsubsec:3.2} for details.
    (An animation of this figure is available.)}
    \label{fig4}
\end{figure}

As depicted in Figures\,\ref{fig4}(a)-(h), the joint observation by the eight channels, i.e., the AIA 94, 335, 211, 193, 171, 131, and 304 \,\AA~and the SUTRI 465\,\AA, unveils the occurrence of cold plasma in the closed loops L1 that is not yet involved in the reconnection. 
The region of the appearance is marked by the green rectangle in Figure\,\ref{fig4}. 
The condensation gradually accumulates and begins to fall along both legs of the loops as CR; see the online animated version of Figure\,\ref{fig4}. 
As the CR is not associated with a flare and open structures, we recognize it as the type II CR, i.e., the quiescent CR along the non-flaring AR closed loops. 

Similar to the analysis of the type I CR in Section\,\ref{subsubsec:3.1}, we measure the type II CR to illustrate the cooling process and the speeds of descent of the condensation. 
Time slices of AIA 304\,\AA~images along the cyan dotted lines KL and MN in Figure\,\ref{fig4}(h) are made, and displayed in Figures\,\ref{fig6}(b1) and (b2). 
There are clear flows in the direction KL, with a velocity of 36 km\,s$^{-1}$ and acceleration of $\sim$58\,m\,s$^{-2}$; see the green dotted lines in Figure\,\ref{fig6}(b1). 
In Figure\,\ref{fig6}(b2), we observe that the condensed material remains stationary near position M (where the condensed material forms) for about 20 minutes before it rains down. This is mainly caused by the evolution of the CR that forms at the top of loops L1, which includes both positions K and M, falls along the direction KL first, and then falls to the solar surface along the direction MN with a speed of 65\,km\,s$^{-1}$ and an acceleration of $\sim$100\,m\,s$^{-2}$, as shown by the green dotted line in Figure\,\ref{fig6}(b2).

The condensation region we choose to calculate the light curves is enclosed by the green rectangles in Figures\,\ref{fig4}(a)-(h). 
As shown in Figures\,\ref{fig7}(b1)-(b7), the normalized light curves for AIA 94, 335, 211, 193, 171, 131, and 304\,\AA~start at 04:40 UT and last for 4\,hr. 
They peak at 04:52, 05:45, 06:59, 07:36, 07:47, 07:53, and 08:14 UT, respectively, marked by the red vertical dotted lines.
This supports the cooling process of the hot plasma in the loops L1.
The coronal plasma thus cools down from $\sim$7.2\,MK (AIA 94\,\AA) to $\sim$2.5\,MK (AIA 335\,\AA) in $\sim$53 minutes, to $\sim$1.9\,MK (AIA 211\,\AA) in $\sim$2.1\,hr, to $\sim$1.5\,MK (AIA 193\,\AA) in $\sim$2.7\,hr, to $\sim$0.9\,MK (AIA 171\,\AA) in $\sim$2.9\,hr, to $\sim$0.6\,MK (AIA 131\,\AA) in $\sim$3\,hr, and to $\sim$0.05\,MK (AIA 304\,\AA) in $\sim$3.4\,hr.
Similar to the type I CR; see Figure\,\ref{fig7}(a4), the SUTRI 465\,\AA~light curve shows a trend similar to the AIA 131\,\AA~light curve; see the blue pluses in Figure\,\ref{fig7}(b6). 
Moreover, nearby the peak of the AIA 94\,\AA~light curve, no peak of the AIA 131\,\AA~light curve is identified, and the AIA 335, 211, 193, 171, 131, and 304\,\AA~light curves decrease evidently.
These results indicate that the plasma in loops L1 is heated up to $\sim$7.2\,MK, the characteristic temperature of AIA 94\,\AA~channel, rather than $\sim$10 MK, the higher characteristic temperature of AIA 131\,\AA~channel.
The peak of the AIA 94\,\AA~light curve at $\sim$07:40 UT represents the emission from plasma with the lower characteristic temperature of AIA 94\,\AA~channel. 

\subsubsection{The type III CR} \label{subsubsec:3.3}
 
\begin{figure}[htbp]
    \centering
    \includegraphics[width=1.0\textwidth]{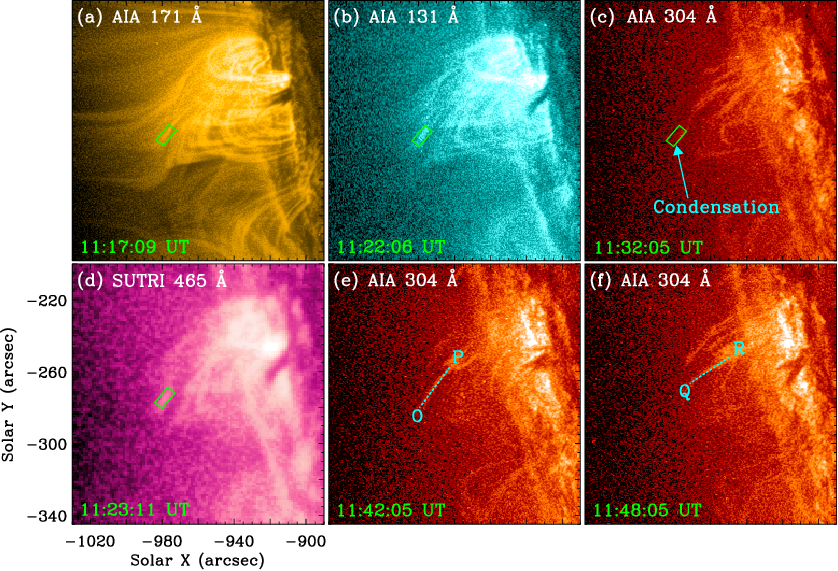}
    \caption{The type III CR. 
    (a) AIA 171, (b) 131 and (c), (e)-(f) 304\,\AA, and (d) SUTRI 465\,\AA~images. 
    The green rectangles in (a)-(d) enclose the region for the light curves in Figure\,\ref{fig7}(c). 
    In (e) and (f), the cyan dotted lines OP and QR mark the positions for the time slices in Figures\,\ref{fig6}(c1)-(c2). 
    An animation of the unannotated AIA images (panels (a)-(c)) is available.
    It covers $\sim$2\,hr starting at 11:00\,UT, with a time cadence of 1\,minute.
    See Section\,\ref{subsubsec:3.3} for details.
    (An animation of this figure is available.)}
    \label{fig5}
\end{figure}

\begin{figure}[htbp]
    \centering
    \includegraphics[width=1.0\textwidth]{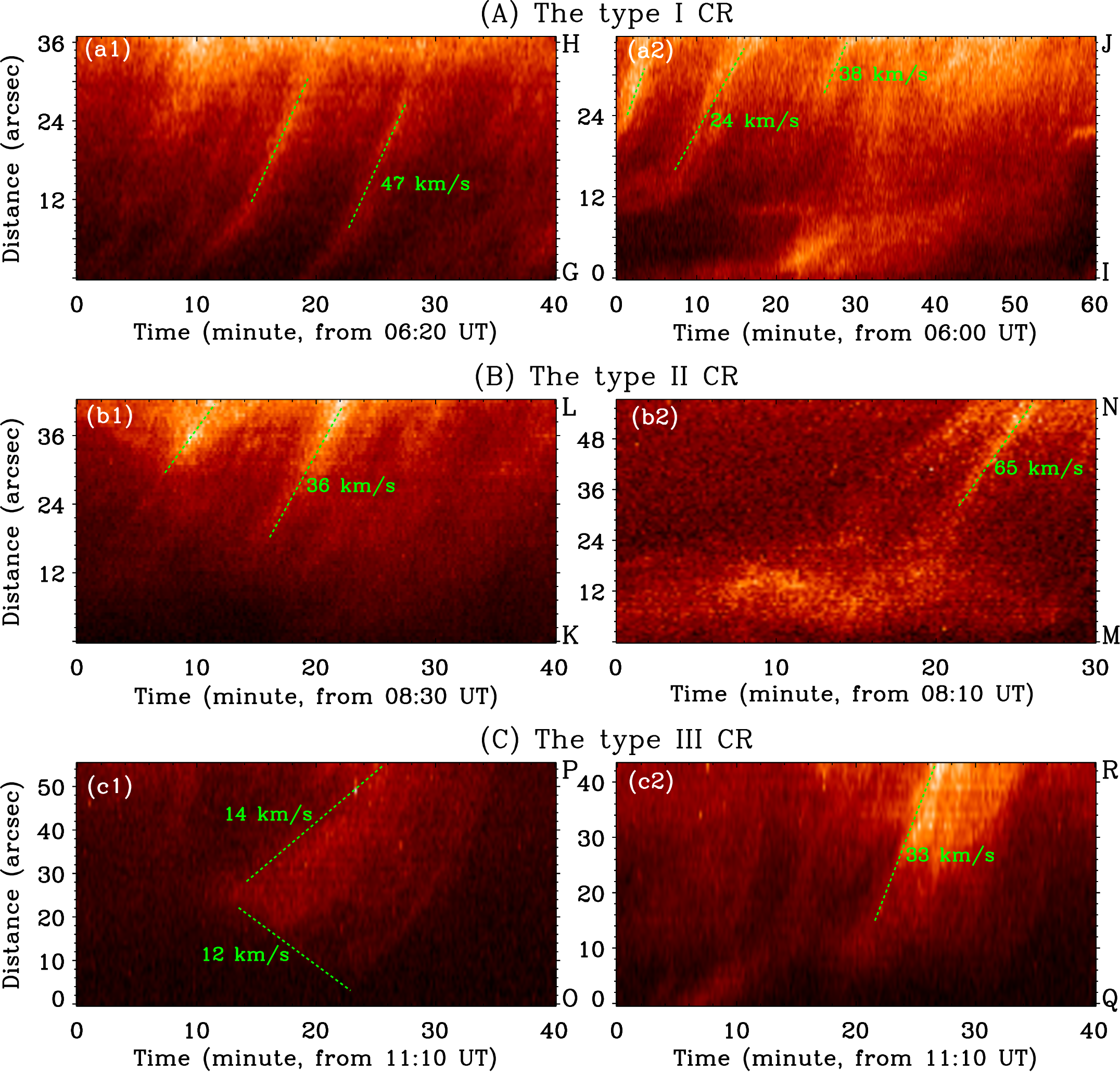}
    \caption{Temporal evolution of three types of CR. Time slices of AIA 304\,\AA~images along the cyan dotted lines GH (a1), IJ (a2), KL (b1), MN (b2), OP (c1), and QR (c2) in Figures\,\ref{fig3}(f), \ref{fig4}(i), and \ref{fig5}(e)-(f), respectively. The green dotted lines outline the motions of CR. See Sections\,\ref{subsubsec:3.1}-\ref{subsubsec:3.3} for details.}
    \label{fig6}
\end{figure}

The interchange reconnection between closed and open structures L1 and L2 here is identical to that previously reported in \citet{2018ApJ...868L..33L}, in which the type III CR forms along the higher-lying open structures.
Similarly, in this study, during the reconnection, condensation forms in the dip of open structures L2; see Figure\,\ref{fig5}(d).
It then falls toward the solar surface along the open structures as CR; see Figures\,\ref{fig5}(e)-(f) and the online animated version of Figure\,\ref{fig5}.
Time slices of AIA 304\,\AA~images are made along the cyan dotted lines OP and QR in Figures\,\ref{fig5}(e)-(f), and illustrated in Figures\,\ref{fig6}(c1) and (c2).
In Figure\,\ref{fig6}(c1), a bidirectional flow of the condensation is observed from $\sim$11:20 UT. 
Here, the downward flow, with the speed and acceleration of 14\,km\,s$^{-1}$ and 90\,m\,s$^{-2}$, respectively, corresponds to the type III CR, and the upward flow may represent the condensations of coronal plasma appearing sequentially from lower- to higher-lying dips \citep{2019ApJ...884...34L}.
Moreover, several downflows are detected in the QR direction in Figure\,\ref{fig6}(c3) with a speed of $\sim$33 km s$^{-1}$ and an acceleration of $\sim$68\,m\,s$^{-2}$. 
Compared with the type I CR, the type III CR is less observed during the reconnection.

\begin{figure}[ht!]
  \centering
  \includegraphics[width=1.0\textwidth]{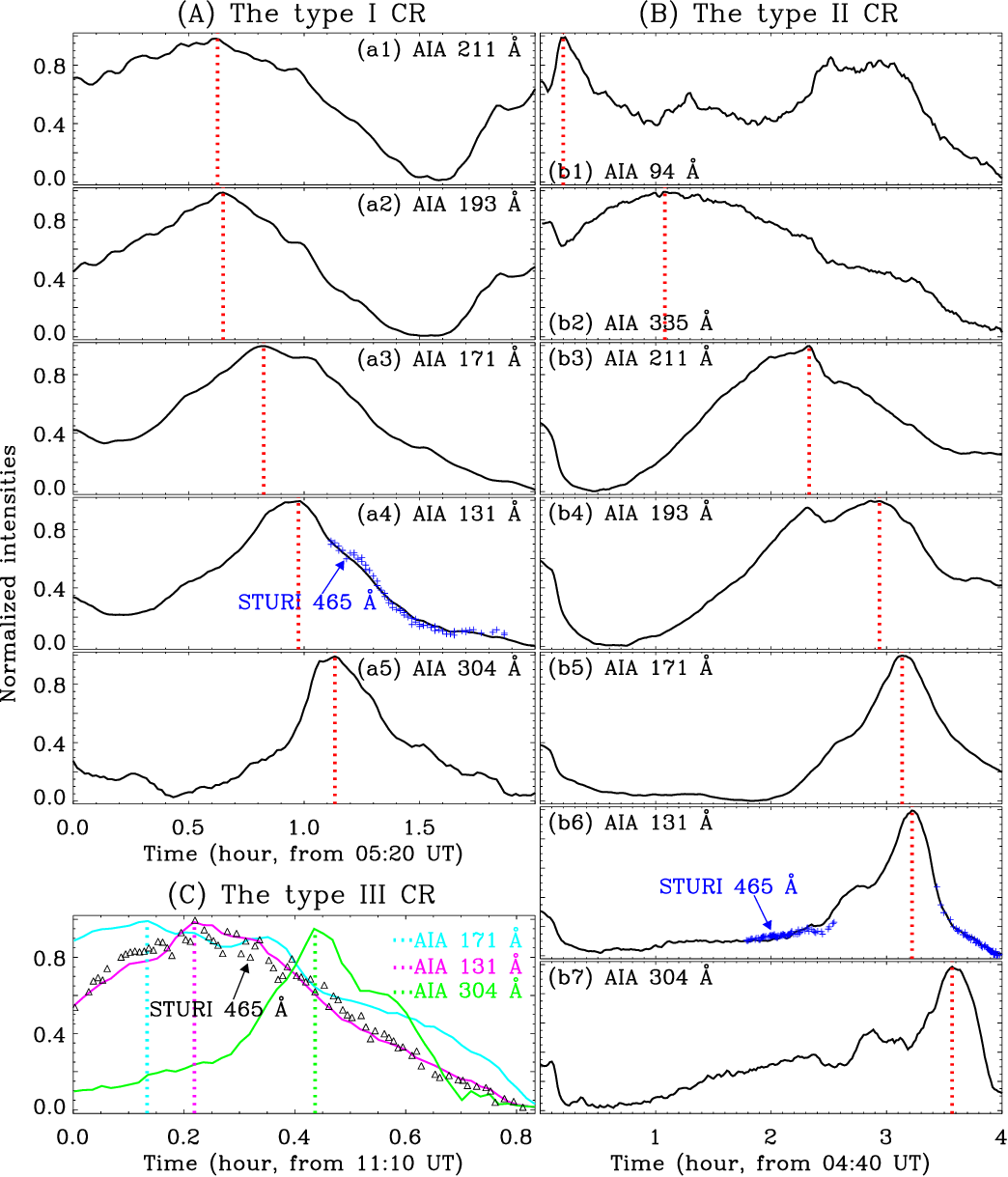}
  \caption{Light curves of three types of CR. (a1)-(a5), (b1)-(b7) and (C) Light curves of the AIA EUV channels in the green rectangles in Figures\,\ref{fig3}-\ref{fig5}, respectively. The red, cyan, pink, and green vertical dotted lines separately mark the peaks of the EUV light curves, and the blue pluses in (a4) and (b6) and the black triangles in (C) denote the light curves of the SUTRI 465\,\AA~channel. See Sections\,\ref{subsubsec:3.1}-\ref{subsubsec:3.3} for details.}
   \label{fig7}
\end{figure}

To better illustrate the plasma cooling process of the type III CR, the light curves of the AIA 171, 131, and 304\,\AA~and SUTRI 465\,\AA~channels are calculated in the region indicated by the green rectangles in Figures\,\ref{fig5}(a)-(d), and displayed in Figure\,\ref{fig7}(d) as the cyan, pink, and green curves and black triangles, respectively. 
No associated structure is observed in the AIA higher temperature, e.g., 94, 335, 211, and 193\,\AA, channels.
The brightening in AIA 131\,\AA~images thus represents the plasma with the lower, rather than the higher, characteristic temperature of the AIA 131\,\AA~channel.
The AIA 171, 131, and 304\,\AA~light curves peak at 11:18, 11:23, and 11:36 UT, respectively, marked by the cyan, pink, and green vertical dotted lines in Figure\,\ref{fig7}(c).
This supports the cooling and condensation process of coronal plasma in the dips of open structures L2.
The coronal plasma thus cools down from $\sim$0.9\,MK, the characteristic temperature of AIA 171\,\AA~channel, to $\sim$0.6\,MK, the lower characteristic temperature of AIA 131\,\AA~channel, in 5 minutes, and then to $\sim$0.05 MK, the characteristic temperature of AIA 304\,\AA~channel, in 13 minutes.
Moreover, the AIA 131\,\AA~and SUTRI 465\,\AA~light curves show similar evolution, consistent with those in Sections\,\ref{subsubsec:3.1}-\ref{subsubsec:3.2}.
This further supports that the AIA 131\,\AA~brightening represents plasma with a temperature of $\sim$0.6 MK.

\section{Summary and Discussions} \label{sec:summary}

In this paper, we utilize EUV images from the SDO and the SUTRI to clearly describe a process of interchange reconnection and the evolution of three types of CR. 
On 2022 October 11, the closed loops L1 and their nearby open structures L2 are observed beyond the southeastern solar limb.
They converge and interact, with the formation of current sheet at the interface.
Reconnection then occurs, and the newly-formed closed loops L3 and open structures L4 recede from the reconnection region.
Before the reconnection, condensation appears in the reconnecting closed loops L1, and falls toward the solar surface as the type II CR.
During the reconnection, condensation occurs in the newly-formed closed loops L3, and rains down along both loop legs as the type I CR.
Moreover, condensation forms in the dips of reconnecting open structures L2, and propagates along the leg of open structures toward the solar surface as the type III CR.
For all three types of CR, the imaging observations and light curves of the AIA and SUTRI EUV channels clearly show the cooling and condensation process of coronal plasma.

\begin{figure}[ht!]
  \centering
  \includegraphics[width=1.0\textwidth]{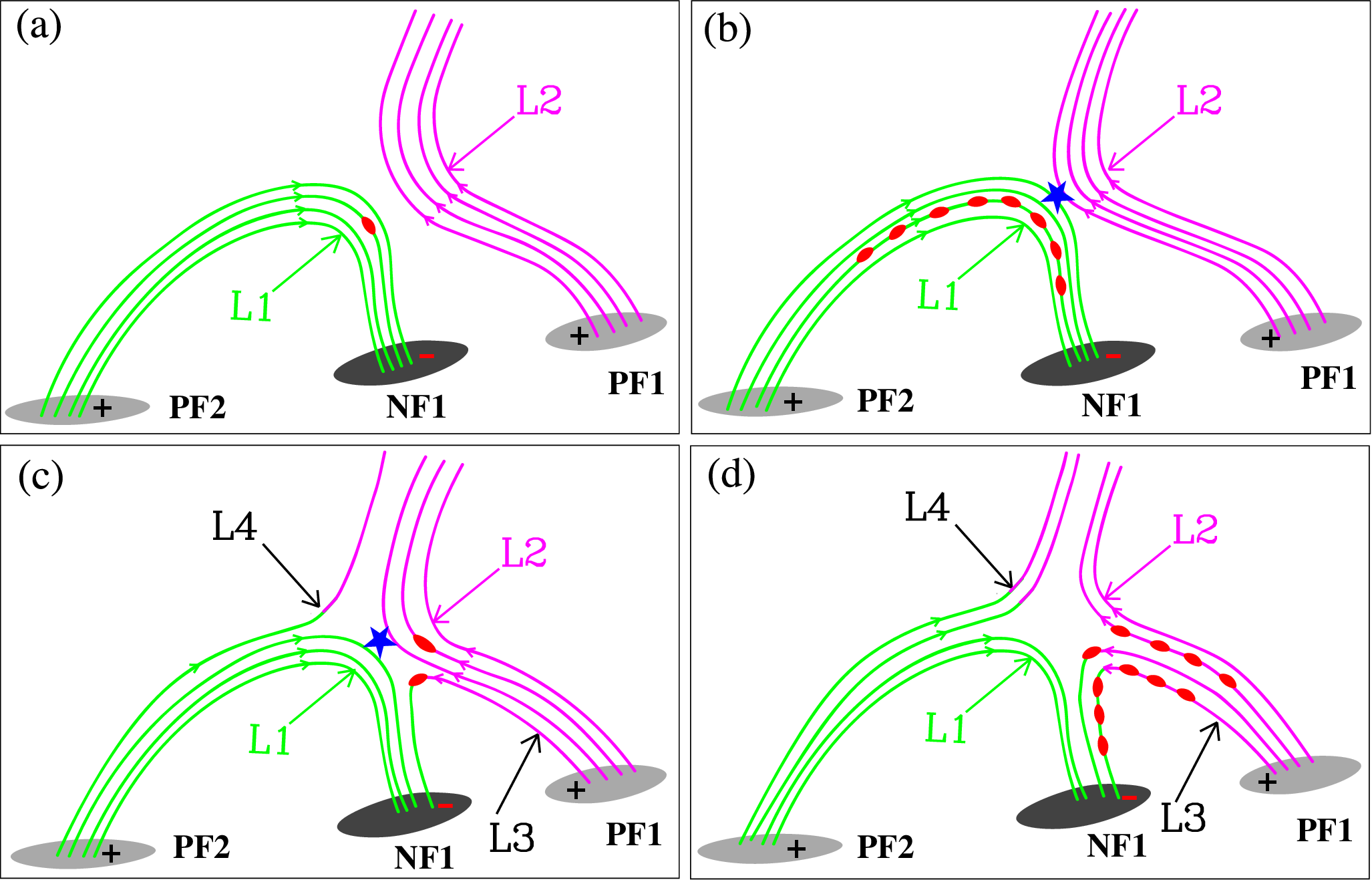}
  \caption{Schematic diagrams of the interchange reconnection and three types of CR. 
  Compared to the observations, the diagrams are rotated $\sim$90$^{\circ}$ clockwise.
  The black, NF1, and gray, PF1 and PF2, ellipses represent the negative and positive magnetic fields, respectively.
  The green and pink lines separately show the closed and open magnetic field lines L1 and L2. 
  The blue stars in (b) and (c) represent the reconnection site. 
  The green-pink lines in (c) and (d) denote the newly-formed closed and open field lines L3 and L4. 
  The red ellipses indicate the condensation. 
  See Section\,\ref{sec:summary} for details.}
  \label{fig8}
\end{figure}

Schematic diagrams, which are rotated $\sim$90$^{\circ}$ clockwise compared to the observations, are presented in Figure\,\ref{fig8} to explain the interchange reconnection and its associated three types of CR. 
PF1 and NF1 separately represent the positive and negative fields of the bipolar region, while PF2 represents the positive field, located to the south of the bipolar region.
Before the reconnection, the heating events at the footpoints of loops L1; see the green lines; lead to the chromospheric evaporation and direct mass ejections into the loops.
Consequently, the loops L1 are rapidly filled with hotter and denser plasma. Due to thermal nonequilibrium, they then cool and condense quickly;
see the red ellipse in Figure\,\ref{fig8}(a).
The condensation falls to the solar surface along both legs of loops L1; see Figure\,\ref{fig8}(b), under the effect of gravity as the type II CR.
Subsequently, the closed loops L1 and open structures L2 (pink lines in Figure\,\ref{fig8}), converge and reconnect (blue star in Figure\,\ref{fig8}(b)).
The newly-formed closed loops L3 and open structures L4 (green-pink lines in Figures\,\ref{fig8}(c) and (d)), then appear and move away from the reconnection region.
The thermal conduction front and/or beamed nonthermal particles, produced during the reconnection, heat the chromospheric material quickly, leading to the chromospheric evaporation and then the greater mass loading of the newly-reconnected loops L3. 
At the top of loops L3, thermal instability takes place.
Cooling and condensation of hot plasma (red ellipse on loops L3 in Figure\,\ref{fig8}(c)), then occur rapidly.
Under the effect of gravity, the condensation moves down to the solar surface along both legs of loops L3 (Figure\,\ref{fig8}(d)) as the type I CR.
Here, it should be noted that, according to the simulation results of \citet{2020ApJ...890..100R}, electron beam heating alone cannot directly result in the formation of CR in loops L3.
During the reconnection, magnetic dips of open structures L2 form upon the reconnection region.
Material around the dips flows into the dips.
This causes an increase in plasma density in the dips. This increase in density triggers thermal instability, which leads to the cooling and condensation of hot plasma within the dips; see the red ellipse on open structures L2 in Figure\,\ref{fig8}(c).
The condensation falls to the solar surface under the effect of gravity along the leg of open structures L2 as the type III CR (Figure\,\ref{fig8}(d)).

Interchange reconnection between closed and open structures is reported.
The converging motions of structures L1 and L2, the appearance and disappearance of current sheet at the interface, and the subsequent formation of newly-reconnected structures L3 and L4 clearly support that interchange reconnection takes place between the closed and open structures L1 and L2.
The moving speeds of structures L1 and L2 (2-3 km\,s$^{-1}$) toward the reconnection region and those of structures L3 and L4 (4-13 km\,s$^{-1}$) away from the reconnection region are consistent with those in \citet{2018ApJ...864L...4L, 2019ApJ...884...34L}, but smaller than those in flares in \citet{2016ApJ...829L..33L} and \citet{2016ApJ...830L...4S}.
The width of current sheet (5.3 Mm) is consistent with those in \citet{2010ApJ...723L..28L} and \citet{2016ApJ...829L..33L}, but larger than those in \citet{2016NatPh..12..847L, 2021ApJ...908..213L, 2022ApJ...935...85L} and \citet{2020A&A...633A.121X}.
The length of current sheet (13\,Mm) is smaller than those in \citet{2016NatPh..12..847L}, but larger than those in \citet{2021ApJ...908..213L, 2022ApJ...935...85L}.
The reconnection rate (0.4) is similar to those in \citet{2018ApJ...858L...4X} and \citet{2021ApJ...908..213L}, but large than those in \citet{2020A&A...633A.121X}.
The temperature (1.3\,MK) and density (1.2$\times$10$^{9}$\,cm$^{-3}$) of current sheet are smaller than those in \citet{2016NatPh..12..847L, 2021ApJ...908..213L, 2022ApJ...935...85L}.
This indicates that less thermal energy may be converted during the reconnection.

Three types of CR are observed in the interchange reconnection.
The speed and acceleration of these CR are consistent with those previously reported \citep[see reviews in][and references therein]{2020PPCF...62a4016A, 2021RAA....21..255L}.
The newly-reconnected loops L3, where the type I CR forms, appear in AIA 211, 193, 171, and 131\,\AA, rather than AIA 94 and 335\,\AA, images; see Section\,\ref{subsubsec:3.1}. 
The plasma in the loops L3 is hence heated up to $\sim$1.9\,MK, the characteristic temperature of AIA 211\,\AA~channel, rather than higher temperatures.
It then cools down, and appears sequentially in the lower temperature, e.g., AIA 193, 171, 131 (SUTRI 465), and 304\,\AA, channels.
The reconnecting loops L1, where the type II CR forms, however, first appears in AIA 94\,\AA~images; see Section\,\ref{subsubsec:3.2}.
Compared with that in the newly-reconnected loops L3, the plasma in the reconnecting loops L1 is thus heated up to a much higher temperature of $\sim$7.2\,MK, the characteristic temperature of AIA 94\,\AA~channel.
It then cools down, and appears sequentially in the lower temperature, e.g., AIA 335, 211, 193, 171, 131 (SUTRI 465), and 304\,\AA, channels.
Comparing with those in the loops L3 and L1 where the types I and II CR form, the coronal plasma with a lower temperature of $\sim$0.9\,MK, the characteristic temperature of AIA 171\,\AA~channel, in the dips of open structures L2 above the reconnection region quickly cools and condenses, forming the type III CR; see Section\,\ref{subsubsec:3.3}. 
It is hardly heated by the reconnection, compared with that of the newly-reconnected loops L3, as the thermal conduction, produced during the reconnection, is inefficient across the field lines.
Even though the formation mechanisms of three types of CR are different, the cooling times of them are comparable.
For example, coronal plasma cools down from $\sim$0.9\,MK (AIA 171\,\AA) to $\sim$0.6\,MK (AIA 131\,\AA) in 10, 6, and 5 minutes, and then to $\sim$0.05\,MK (AIA 304\,\AA) in 9, 21, and 13 minutes for the types I, II, and III CR, respectively.

The types I and III CR are commonly associated with reconnection \citep{2018ApJ...864L...4L, 2019ApJ...884...34L,li2020relation, 2021RAA....21..255L, 2020PPCF...62a4016A}.
In this study, more type I CR and a few type III CR are observed during the interchange reconnection.
This is different from \citet{2018ApJ...864L...4L, 2019ApJ...884...34L, li2020relation} in which the type III CR, rather than the type I CR, is detected.
Comparing the interchange reconnection event here with those in \citet{2018ApJ...864L...4L, 2019ApJ...884...34L, li2020relation}, we may answer the question that under what conditions the types I and III CR form during the interchange reconnection.
Compared with the reconnection events in \citet{2018ApJ...864L...4L, 2019ApJ...884...34L, li2020relation}, in this study, similar converging speeds of reconnecting structures toward the reconnection region and moving speeds of newly-reconnected structures away from the reconnection region are observed.
However, the reconnecting closed loops L1, where the type II CR that always occurs in the non-flaring AR closed loops forms, and the newly-reconnected closed loops L3, where the type I CR takes place, are clearly identified.
Moreover, the current sheet evidently appears at the interface between the reconnecting open and closed structures.
These observations indicate that more magnetic flux participates in the interchange reconnection region with a similar inflowing speed in this study.
More magnetic energy is thus converted into more other, e.g., thermal and kinetic, energy in a similar or even short time, showing a larger reconnection rate.
Furthermore, we calculate the reconnection rate, the ratio of the width and the length of the current sheet, in the interchange reconnection events in \citet{2018ApJ...864L...4L} and \citet{li2020relation}, and get similar values of $\sim$0.2, half of the reconnection rate (0.4) in this study.
We hence propose that the reconnection rate plays an essential role in the formation of types I and III CR during the interchange reconnection.
The type I (III) CR tends to form in the interchange reconnection between open and closed structures with a larger (smaller) reconnection rate.

\subsection{Acknowledgments}

The authors thank the referee for helpful comments that led to improvements in the munascript. We are indebted to the SDO and SUTRI teams for providing the data.
This work is supported by the National Key R\&D Programs of China (2022YFF0503002 (2022YFF0503000)), the National Natural Science Foundations of China (12073042, U2031109, 12350004 and 12273061), China Postdoctoral Science Foundation (No. 2021M700246), the Key Research Program of Frontier Sciences (ZDBS-LY-SLH013) and the Strategic Priority Research Program (No. XDB 41000000) of Chinese Academy of Sciences, and Yunnan Academician Workstation of Wang Jingxiu (No. 202005AF150025).
AIA images are courtesy of NASA/SDO and the AIA, EVE, and HMI science teams. 
SUTRI, on the other hand, is a collaborative project undertaken by the National Astronomical Observatories of Chinese Academy of Sciences, Peking University, Tongji University, Xi'an Institute of Optics and Precision Mechanics of Chinese Academy of Sciences, and the Innovation Academy for Microsatellites of Chinese Academy of Sciences.

\bibliography{sample631}{}
\bibliographystyle{aasjournal}



\end{document}